\documentclass[aps,prl,superscriptaddress,nofootinbib,twocolumn]{revtex4-2}
\usepackage{graphicx}  
\usepackage{bm}        
\usepackage{amssymb}   
\usepackage{amsmath}
\usepackage{color}
\usepackage{xcolor}
\usepackage{ulem}
\newcommand{\el}[1]{\textcolor{black}{#1}}

\newcommand{\er}[1]{\textcolor{black}{#1}}

\begin{document}

\title{Non-Markovian feedback control and acausality: an experimental study}
\author{Maxime Debiossac}
\email{maxime.debiossac@univie.ac.at}
\affiliation{University of Vienna, Faculty of Physics, VCQ, Boltzmanngasse 5, A-1090 Vienna, Austria} 
\author{Martin Luc Rosinberg}
\affiliation{LPTMC, CNRS-UMR 7600, Sorbonne Universit\' e, 4 place Jussieu, 75252 Paris Cedex 05, France}
\author{Eric Lutz}
\affiliation{Institute for Theoretical Physics I, University of Stuttgart, D-70550 Stuttgart, Germany}
\author{Nikolai Kiesel}
\affiliation{University of Vienna, Faculty of Physics, VCQ, Boltzmanngasse 5, A-1090 Vienna, Austria}

\begin{abstract} 

Causality is an important assumption \el{underlying} nonequilibrium generalizations of the second law of thermodynamics known as fluctuation relations. We here experimentally study the nonequilibrium statistical properties of the work and of the entropy production  for an optically trapped, underdamped nanoparticle  continuously subjected  to a time-delayed feedback control. Whereas the non-Markovian  feedback  depends on the past position of the particle  for a forward trajectory, it depends on its future position  for a time-reversed path, and is therefore acausal. In the steady-state regime, we show that  the corresponding fluctuation relations in the long-time limit exhibit  a clear signature of this acausality, even though the time-reversed dynamics is not  physically realizable.
\end{abstract}
\maketitle

There is a direct relationship between heat dissipation and  irreversibility of a thermodynamic process, as expressed by the breaking of time-reversal symmetry. Consider, for example, a classical system in contact with a thermal bath, such as a Brownian particle driven arbitrarily far from equilibrium by an external perturbation. The heat exchanged with the bath  along a particular stochastic trajectory (starting from a given initial state) can  then be expressed as the logratio  of the probabilities of the trajectory and  of the corresponding time-reversed path, with the time-reversed  protocol \cite{S2012}.
This fundamental property,  referred to as microscopic reversibility~\cite{C1999}, ensures  the thermodynamic consistency of the dynamics at the level of fluctuating trajectories. It is at the origin of most  fluctuation relations, which are the cornerstones of our modern understanding of out-of-equilibrium processes~\cite{E2002,C1999,J2011,S2012,C2017}.

Matters are more complicated when information is  extracted from the system  to regulate or modify its state via feedback control { \cite{PHS2015,HV2010,SU2012,TSUMS2010,KMSP2014,rol14,rib19,ric21}}. Feedback operation is widespread in many  biological systems and technological applications \cite{ast10,BR2000,A2010,bec21}. Non-Markovian effects induced by the unavoidable time lag between  signal detection and  control action then raise a fundamental question:  how  should  the time-reversed process, associated with microscopic reversibility, be defined?  Should  the reversed dynamics be implemented without measurement and feedback, which is the solution advocated for Szil\'ard-type engines~{\cite{TSUMS2010,HV2010,SU2012}}? Or should the two still be present? In the latter case, microscopic reversibility  is modified, as the fluctuating heat  is no longer  odd under time reversal~\cite{MR2014,RMT2015,RTM2017}.  Since feedback action depends on the future state of the system  for the backward process, causality is violated (Fig.~\ref{fig:Trajectory}). 
Still, this strategy is claimed to provide a consistent  description of the stochastic thermodynamics~\cite{MR2014,RMT2015,RTM2017}. But there is a  price to pay:  the  backward process is no longer  physically realizable. 
\begin{figure}[t]	\includegraphics[width=0.7\linewidth]{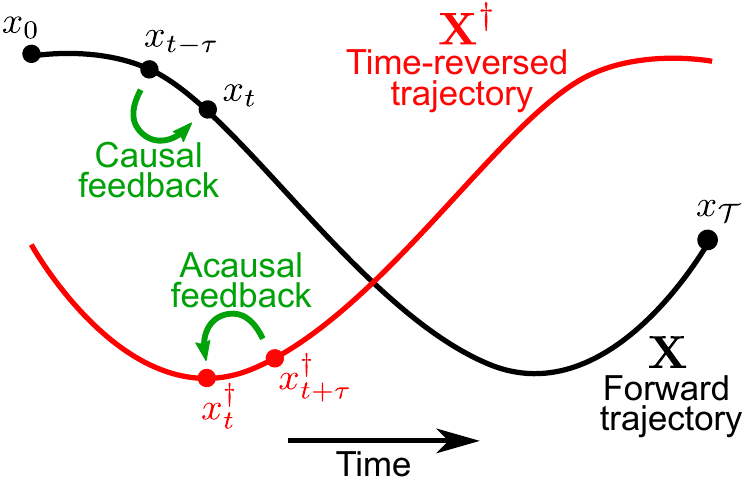}
	\caption{Schematic representation of a  path ${\bf X}=\{x_t\}_{0\le t\le {\cal T}}$ and  its time-reversal ${\bf X}^\dag=\{x_{{\cal T}-t}\}_{0\le t\le {\cal T}}$. While the time-delayed feedback control depends on the past position of the particle $x_{t-\tau}$ (causal feedback) for the forward trajectory, it depends on the future position of the particle $x^\dag_{t+\tau}$ (acausal feedback) for the time-reversed path. 
\label{fig:Trajectory}} 
\end{figure}
 
In this paper, we report the first experimental evidence  that the acausal backward dynamics is a useful tool that allows us to predict properties of real physical systems. 
To this end, we use a setup consisting of a levitated nanoparticle trapped in a harmonic potential and subjected to  a position-dependent time-delayed feedback~\cite{GFHKA2016,DGALK2020}. Owing to the continuous feedback, the system settles in a nonequilibrium steady state  where heat is permanently exchanged with the bath.
We then record the  fluctuations of two trajectory observables: the work  ${\cal W}_{{\cal T}}$ done by the feedback force  within a  time window of duration ${\cal T}$ and the corresponding  entropy production  $\Sigma _{\cal T}$. Whereas the expectation values of these  quantities are identical, their fluctuations far away from the mean may differ, due to rare events associated with temporal boundary terms. Their statistics, moreover, depend on the delay \cite{MR2014,RMT2015,RTM2017}.  A remarkable conjecture is that  $\Sigma _{\cal T}$ satisfies  asymptotically the fluctuation relation~\cite{MR2014,RMT2015,RTM2017} 
\begin{equation}
\label{EqIFT}
\langle e^{-\Sigma_{{\cal T}}}\rangle\sim e^{\dot S_J {\cal T}}\ , \:\: ({\cal T} \to \infty)\ ,
\end{equation}
where $ \dot S_J$ is a  ``Jacobian" contribution induced by the breaking of causality in the backward process (the theoretical  analysis is summarized in the Supplemental Material~\cite{SI}). The rate $\dot S_J$ is  an upper bound to the extracted work~\cite{RMT2015}  and  may  be viewed as an entropic cost of the non-Markovian feedback. When the  acausal backward dynamics  allows for the existence of a stationary state, the work  ${\cal W}_{{\cal T}}$ is predicted to similarly obey $\langle e^{-\beta {\cal W}_{{\cal T}}}\rangle {\sim} e^{\dot S_J {\cal T}}$, with $\beta=1/(k_BT)$ and $T$  the  bath temperature)~\cite{RTM2017,SI}. 

In the following, we experimentally check the validity of these two asymptotic  fluctuation relations. A confirmation of these conjectures would open the door to an experimental determination of  the acausal contribution  $\dot S_J$, whose explicit expression for general non-Markovian Langevin systems is out of reach~\cite{note1}.  However, estimating an  exponential average from experimental time series is a daunting task, as illustrated by the computation of the equilibrium free energy from the Jarzynski equality~\cite{J1997}. The main problem is that the average is dominated by very rare realizations~\cite{J2006}, which by definition are hard to get  in experiments~\cite{note4}. Obtaining the genuine asymptotic behavior is even more  challenging since it becomes exponentially more unlikely to observe fluctuations away from the mean as ${\cal T}$ increases. This is the major difficulty we have to address and  this forces us to perform a delicate statistical analysis of the behavior of  $\langle e^{-\Sigma_{{\cal T}}}\rangle$ and $\langle e^{-\beta {\cal W}_{{\cal T}}}\rangle$  as a function of  the time ${\cal T}$.

\begin{figure}[t]	\includegraphics[width=0.86\linewidth]{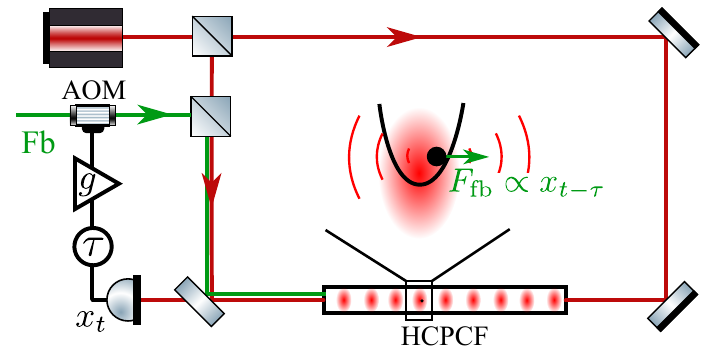}
	\caption{Experimental setup. A nanoparticle is levitated in a harmonic optical trap formed by two counter-propagating beams (red), inside a hollow-core photonic crystal fiber. The nanoparticle is  subjected to a delayed feedback force $F_\text{fb}\propto x_{t-\tau}$ implemented with an acousto-optic modulator using radiation pressure from an additional laser beam (green). The feedback loop is characterized by a gain $g$ and time delay $\tau$.
	\label{fig:fluctuations}} 
\end{figure}
 
\textit{Experimental setup.} We consider a levitated nanoparticle (295 nm diameter)  trapped at an intensity maximum of a standing wave formed by two counterpropagating laser beams ($\lambda=1064$ nm) inside a hollow-core photonic crystal fiber~\cite{GFHKA2016,DGALK2020} (Fig.~\ref{fig:fluctuations}). The particle oscillates in the harmonic trap with a resonance frequency $f_0=\Omega_0/2\pi=297.7$ kHz and is damped by the surrounding gas at temperature $T=293$ K with a damping rate $\Gamma_0/2\pi=5.93$  kHz. The quality factor of the oscillator is $Q_0=\Omega_0/\Gamma_0=50.2$. The particle motion along the fiber axis ($x$-axis) is detected by interferometric readout of the light scattered by the particle, with a position sensitivity of 2 pm/$\sqrt{\text{Hz}}$~\cite{GFHKA2016}. A variable delay $\tau$ is then added to the position signal  using a field-programmable gate array and a feedback force $F_\text{fb}\propto x_{t-\tau}$ is applied to the nanoparticle via radiation pressure. The feedback loop has a variable  gain $g$ and an internal minimum delay of $3~\mu$s mainly due to the acousto-optic modulator and the band pass filter of the feedback loop.
For each of the 19  chosen delays $\tau$, we record a long trajectory of duration $\mathcal{T}_\text{tot}=1000\:$s with a sampling rate of $5$ MHz. This amounts to 5 hours total data acquisition time, which is short enough to keep experimental drifts small (estimated relative uncertainties:  ${\Delta \Gamma_0}/{\Gamma_0} \leq 2\%$ and ${\Delta g}/{g} \leq 5\%$). We also  account for the change of sensitivity of the detector by normalizing each raw value of the particle position by the laser power. The  data is  filtered with a $\pm 150$ kHz bandwidth digital filter around $\Omega_0$ to eliminate low frequency technical noise.

For small displacements,  the motion of the particle is  described by an underdamped Langevin equation \cite{ris89},
\begin{equation}
\ddot{x}_t+ \Gamma_0 \dot{x}_t+ \Omega_0^2 x_t- g \Gamma_0 \Omega_0 x_{t-\tau}=\sqrt{\frac{2\Gamma_0 k_BT}{m}}\xi_t,
\label{eq:langevin}
\end{equation}  
where  $m$ is the mass and $\xi(t)$ is the Gaussian thermal noise, delta-correlated in time with variance $1$.  The feedback force is $F_{\text{fb}}= -g (m \Gamma_0 \Omega_0) x_{t-\tau}$ with  $g>0$. Hereafter, we shall use $\Omega_0^{-1}$ and $x_0=(1/\Omega_0^2)(2\Gamma_0 k_BT/m)^{1/2}$ as units of time and position, respectively~\cite{MR2014,RMT2015,RTM2017}. 
The dynamics of the particle is then  characterized by the dimensionless parameters $(g,Q_0,\tau)$~\cite{note5}. We further express the observation time  ${\cal T}$  in units of  $Q_0$, that is, as a multiple of the relaxation time $\Gamma_0^{-1}$.

\begin{figure}[t]
\begin{center}
\includegraphics[width=0.98\linewidth]{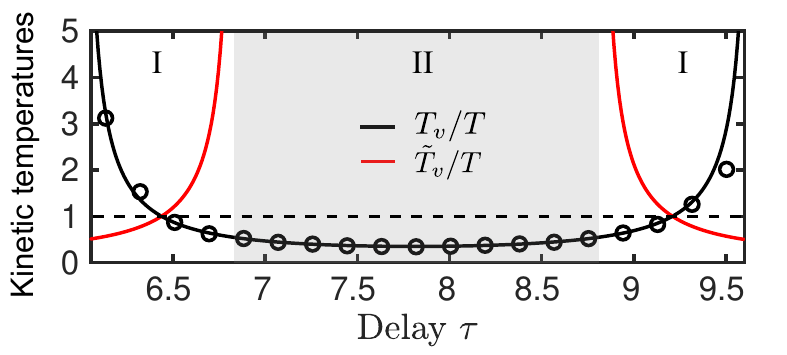}
\caption{\label{fig:temperature} Steady-state kinetic temperatures versus time delay in the second stability region of the oscillator (in units of $\Omega_0^{-1}$).  The black  solid  line is the theoretical prediction for the (causal) temperature ratio $T_v/T$  and  symbols are  data obtained by averaging  $v^2_t$  over a trajectory of length $1000 Q_0$. The red solid line is the  prediction for the (acausal) temperature ratio ${\widetilde T}_v/T$ in the region (labelled I) where the acausal dynamics admits a stationary solution.}
\end{center}
\end{figure}

\begin{figure*}[t]
\begin{center}
\includegraphics[width=0.76\linewidth]{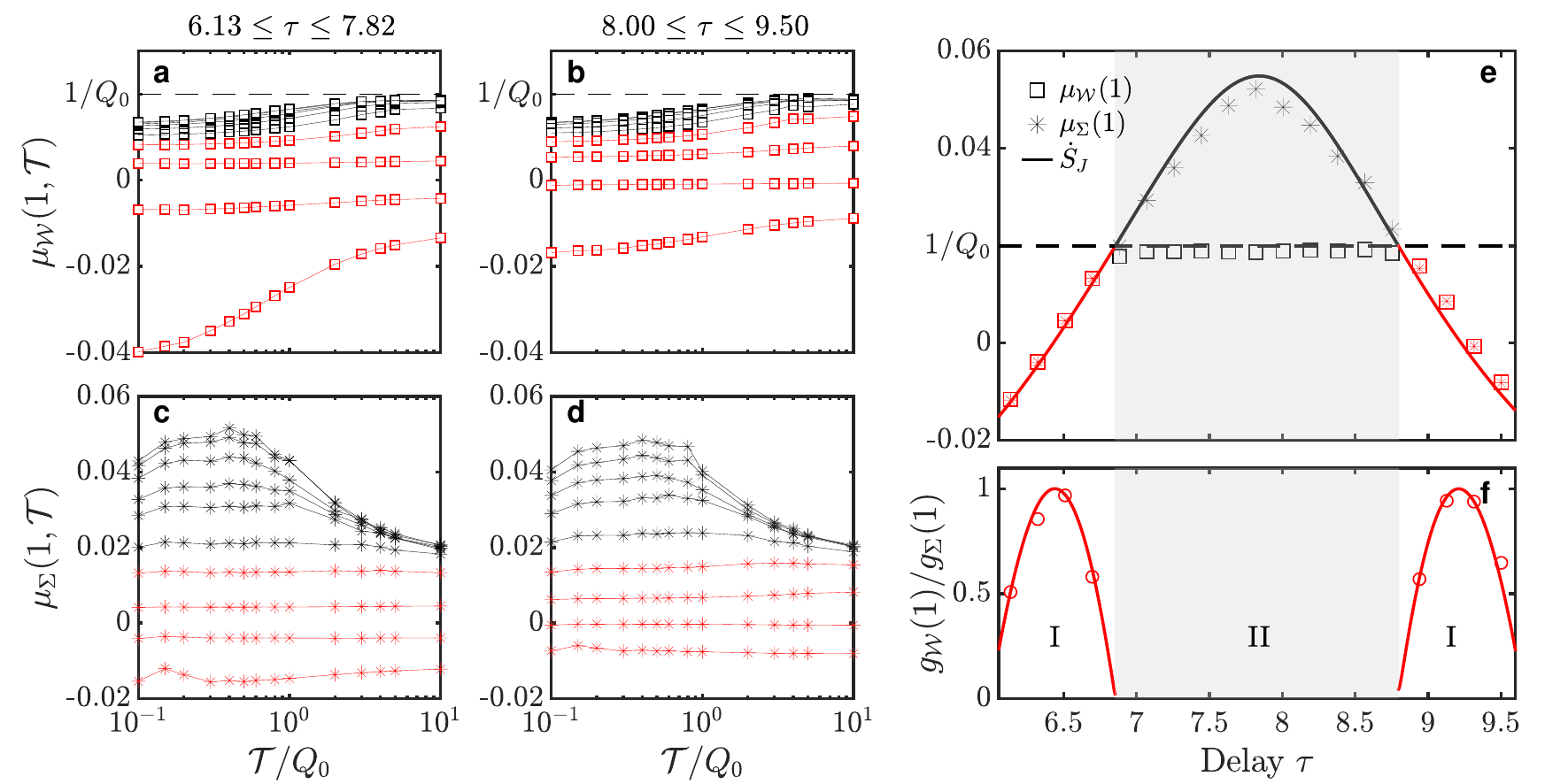}
\caption{\label{fig:SJ} (a-d) Estimates of $\mu_{W} (1,{\cal T})$ (squares) and $ \mu_{\Sigma}(1,{\cal T})$ (stars) as a function of the observation time ${\cal T}$ (in unit of $\Omega_0^{-1}$), for various delays $\tau$. Panels  (a) and (c): from bottom to top, $\tau=6.13,6.32,6.51, 6.70$ (red) and $\tau=6.88,7.07,7.26,7.44,7.63,7.82$ (black). Panels (b) and (d):  from top to bottom, $\tau=8.00,8.19,8.38,8.57,8.75$ (black) and $\tau=8.94,9.13,9.31,9.50$ (red). Solid lines are a guide to the eye. e) Values of $\mu_\Sigma(1)$ (stars) and  $\mu_\mathcal{W}(1)$ (squares) deduced from the  data plotted in panels (a)-(d). The solid  line is the theoretical expression of the acausal contribution $\dot{S}_J$~\cite{RMT2015,SI}. In region I (red symbols), the acausal dynamics has a stationary solution and $\mu_{\Sigma}(1)=\mu_W(1)= \dot S_J$. In region II (black symbols), $\mu_W(1)= 1/Q_0$ and  $\mu_{\Sigma}(1)= \dot S_J$. f) Ratio of the pre-exponential factors $g_{W}(1)/g_{\Sigma}(1)$ in region I. The red solid line is the  conjectured formula (\ref{Eqratio}). \er{Error bars are discussed in the Supplemental Material~\cite{SI}.}}
\end{center}
\end{figure*}

The feedback-controlled oscillator has a complex dynamical behavior and may exhibit multistability~\cite{RMT2015}. We choose a feedback gain $g=2.4$ so as to probe the two regimes that  differentiate the fluctuations for  ${\cal W}_{{\cal T}}$ and $\Sigma_{{\cal T}}$ (Fig.~\ref{fig:SJ}). We moreover select values of $\tau$ between 6.13 and 9.50 (i.e. between $3.28\:\mu$s and $5.08\:\mu $s) that correspond to the second stability region~\cite{SI}. 
The viscous relaxation time  ($\Gamma_0^{-1}\approx 27\mu$s) is much larger than the considered delays,  ensuring the efficiency of the feedback loop. This is quantified by the  kinetic  temperature  of the particle, defined as $T_v=(2T/Q_0)\langle v^2_t\rangle$ in reduced units, which  determines the average heat flow into the environment, $\langle \beta \dot Q\rangle=(1/Q_0)(T_v/T-1)$~\cite{RMT2015,DGALK2020}.  As seen in Fig.~\ref{fig:temperature}, the experimental values of $T_v$  computed from the mean-square velocity (black symbols) are in excellent agreement with the theoretical predictions  (black lines)~\cite{RMT2015}.
In particular, the feedback cooling regime, $T_v/T<1$, and thus $\langle \dot Q\rangle<0$, is achieved  for $6.44\lesssim\tau\lesssim9.21$.

 \textit{Asymptotic fluctuation relations for  work and entropy production.}
The  stochastic work ${\cal W}_{{\cal T}}$  performed by the  feedback force along a trajectory of duration ${\cal T}$  is 
\begin{equation}
\label{EqWXY}
\beta {\cal W}_{{\cal T}}=\frac{2g}{Q_0^2}\int_0^{{\cal T}} dt \: x_{t-\tau}\circ v_{t} \ ,
\end{equation}
 where the integral is interpreted with the Stratonovich prescription~\cite{S2010}. The corresponding  entropy production (EP) is  defined as~\cite{SI}
  \begin{equation}
\label{EqSigmaXY}
\Sigma_{{\cal T}}= \beta {\cal Q}_{{\cal T}}+\ln \frac{p(x_0,v_0)}{p(x_{{\cal T}},v_{{\cal T}})}\ ,
\end{equation} 
 where ${\cal Q}_{{\cal T}}=(2/Q_0)\int_0^{{\cal T}} dt\: [v_{t}/Q_0-\xi(t)] \circ v_{t}$ is  the heat dissipated into the environment~\cite{S2010} and $p(x,v)$ is the stationary probability distribution.  Note that $\langle e^{-\Sigma_{{\cal T}}}\rangle\ne 1$ and  that $\langle \Sigma_{{\cal T}}\rangle <0$  in the cooling regime, which may look as a violation of the second law. But $\Sigma_{{\cal T}}$ is  just  the ``apparent"   stochastic EP which an observer unaware of the existence of the feedback loop would  naively regard as the total  EP~\cite{S2005}. The important point is that $\Sigma_{{\cal T}}$ is an experimentally accessible quantity.

To simplify the  analysis,  we  rewrite Eq.~(\ref{EqIFT})  as  $\mu_{\Sigma}(1)=\dot S_J$ where $\mu_{\Sigma}(\lambda)=  \lim_{{\cal T}\to \infty}(1/{\cal T})\ln \langle e^{-\lambda \Sigma_{{\cal T}}}\rangle$ is the  scaled cumulant generating function~\cite{T2009}. Likewise, $\mu_{W}(1)=\dot S_J$ when the acausal dynamics converges to a stationary state.
Correctly estimating these two quantities from experiments requires a large amount of data. We use a block-averaging approach, dividing the full time series $\{x_t\}_{0 \le t\le {\cal T}_\text{tot}}$ into $N$ blocks of length ${\cal T}$ \cite{SI}. We then compute the  observables  in each block, and  approximate $ \langle e^{-\Sigma_{{\cal T}}}\rangle$  and $ \langle e^{-\beta {\cal W}_{{\cal T}}}\rangle$ with the averages over the $N$ blocks.  Obviously, a large value of ${\cal T}$ implies a small ensemble size $N$, resulting in  a  poor estimation of the empirical averages. There is no perfect solution to this conundrum~\cite{RAT2015}, but  we found preferable  to use a large value of $N$  and   study the two quantities  $\mu_{\Sigma}(1,{\cal T})=(1/{\cal T}) \ln \langle e^{-\Sigma_{{\cal T}}}\rangle$  and $ \mu_{W}(1,{\cal T})=(1/{\cal T}) \ln \langle e^{-\beta {\cal W}_{{\cal T}}}\rangle$ as a function of  the observation time ${\cal T}$.
\er{We then analyze their large-time limits by fitting them for finite $\mathcal{T}$ and extrapolating the results to $\mathcal{T}\to\infty$ (see Ref.~\cite{hil17} for a similar procedure).}

\textit{Results.} The estimates of $\mu_{\Sigma}(1,{\cal T})$ and $\mu_{W}(1,{\cal T})$ are presented in   Figs.~\ref{fig:SJ}a,c for $6.13\le \tau\le 7.82$   and  Figs.~\ref{fig:SJ}b,d for $8.00\le \tau \le 9.50$. The ensemble size is $N = 5 \cdot 10^6$ for ${\cal T}\le  5 Q_0$ and  $N = 3 \cdot 10^6$ for ${\cal T}=10 Q_0$. 
\er{We} observe that $\mu_{\Sigma}(1,{\cal T})$ and $\mu_{W}(1,{\cal T})$ behave differently with ${\cal T}$. This is due to contributions  of the  boundary term in $\Sigma_{{\cal T}}$. More interesting is the fact that two  regimes, hereafter labelled I and II, can be distinguished as a function of  the delay:  In regime I (red symbols in  Figs.~\ref{fig:SJ}a-d),  both $\mu_{\Sigma}(1,{\cal T})$  and $\mu_{W} (1,{\cal T})$ vary more or less monotonically with ${\cal T}$ and the  values for the largest observation time depend on $\tau$. By contrast,  in regime II (black symbols),  these  values are  independent  of $\tau$ and very close to $1/Q_0\approx 0.02$.  Moreover, $\mu_{\Sigma}(1,{\cal T})$ first increases with ${\cal T}$, then reaches a plateau, and ultimately decreases to $1/Q_0$. We will argue below that this decrease is induced by the finite statistics of rare events.

The existence of two regimes  suggests an intriguing connection  with  the   acausal dynamics obtained by changing $x_{t-\tau}$  into $x_{t+\tau}$ in  Eq.~(\ref{eq:langevin}). Indeed, this dynamics admits a stationary solution for $6.00\lesssim \tau\lesssim 6.85$ and $8.80\lesssim \tau\lesssim 9.66$~\cite{SI}. This ``acausal" stationary state is characterized by finite values of $\langle x^2_t\rangle$ and $\langle v^2_t\rangle$, and thus well-defined ``acausal" configurational and kinetic temperatures $\widetilde T_x$ and $\widetilde T_v$~\cite{RTM2017,SI}. This is illustrated in  Fig.~\ref{fig:temperature} that shows  the variations of  ${\widetilde T}_v$ with $\tau$.

\er{In this regime I,} theory predicts that  $\mu_{\Sigma}(1)=\mu_W(1)=\dot S_J$~\cite{RTM2017}. To experimentally check these conjectures, we write  $ \langle e^{-\Sigma_{{\cal T}}}\rangle {\sim} g_{\Sigma}(1)\,e^{\mu_{\Sigma}(1){\cal T}}$ and $ \langle e^{-{\cal W}_{{\cal T}}}\rangle {\sim} g_{W}(1)\,e^{\mu_{W}(1){\cal T}}$ for ${\cal T}$ sufficiently large~\cite{note3}, and we fit $\mu_{\Sigma}(1,{\cal T})$ and $\mu_{W}(1,{\cal T})$ with   $a+b/{\cal T}+c/{\cal T}^2$ in the  range $1\le {\cal T}/Q_0<10$~\cite{SI}. We  then identify $\mu_{\Sigma}(1)$ [resp.~$\mu_{W}(1)$] with $a$ and the pre-exponential factor $g_{\Sigma}(1)$ [resp.~$g_{W}(1)$] with $e^b$. As shown in Fig.~\ref{fig:SJ}e, the agreement with the theoretical expression of  the acausal contribution $\dot S_J$ is excellent for both quantities. Remarkably, the ratio of the prefactors (Fig.~\ref{fig:SJ}f) is  well represented by the \er{conjectured} formula~\cite{RTM2017},
\begin{equation}
\label{Eqratio}
\!\frac{g_{W}(1)}{g_{\Sigma}(1)}\!=\!\frac{T^2}{\sqrt{[T(T_x+\widetilde T_x)-T_x\widetilde T_x][T(T_v+\widetilde T_v)-T_v\widetilde T_v]}}\ ,
\end{equation}
which depends explicitly on  the "acausal" temperatures $\widetilde T_x$ and $\widetilde T_v$.
\er{This confirms that  the acausal process can  be used to describe experimentally accessible quantities}.

We next determine the asymptotic value of $\mu_{W}(1,{\cal T})$  in regime II. In this case, the  acausal dynamics does not admit a stationary solution, but it can be shown that  $\mu_W(1)=1/Q_0$ (i.e., $\mu_W(1)=\Gamma_0/m$ in real units), independently on the value of $\tau$~\cite{RTM2017,SI}. By fitting  $\mu_{W}(1,{\cal T})$ by the same expression as in regime I,  this is again very well confirmed by the  data (Fig.~\ref{fig:SJ}e).

Verifying that $\mu_{\Sigma}(1)=\dot S_J$  in regime II requires a more delicate analysis. The fact that the intermediate plateau for $\mu_{\Sigma}(1,{\cal T})$ progressively disappears  as $\tau$ is varied from $6.88$ to $7.82$ in Fig.~\ref{fig:SJ}c,  then reappears as $\tau$ is varied from $8$ to $8.75$ in Fig.~\ref{fig:SJ}d, before it disappears again, suggests that the rare events associated with the  boundary term in $\Sigma_{{\cal T}}$ are not correctly sampled. Although this term does not grow with time, it may fluctuate to order ${\cal T}$ and  contribute to $ \langle e^{-\Sigma_{{\cal T}}}\rangle$, but  the probability of such event gets smaller and smaller as ${\cal T}$ increases.  To test the hypothesis that the fall-off of $\mu_{\Sigma}(1,{\cal T})$ for ${\cal T}/Q_0\gtrsim 1$ is artificial, we evaluate  $ \langle e^{-\Sigma_{{\cal T}}}\rangle$,  excluding from the empirical average  events with a  probability $P(\Sigma_{{\cal T}}=\sigma {\cal T})$  smaller than some threshold $\theta$ (Fig.~\ref{Fig: convergence}a).  
\begin{figure}[t]
\begin{center}
\includegraphics[width=1\linewidth]{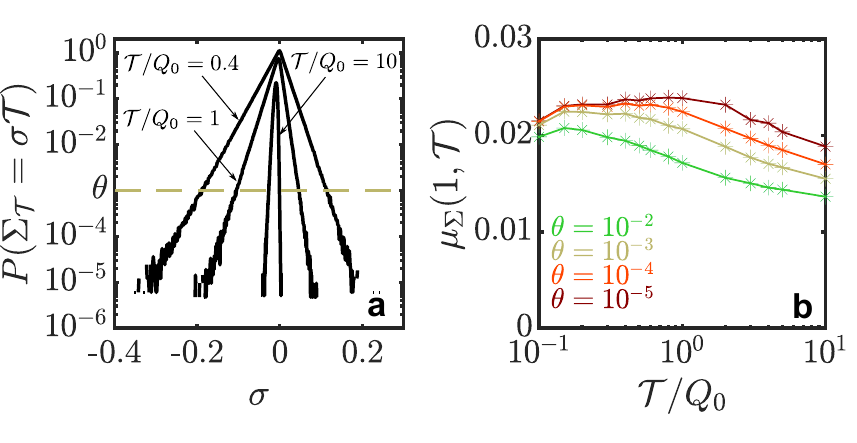}
\caption{\label{Fig: convergence} (a) Experimental distributions of the entropy production $P(\Sigma_{\cal T}= \sigma {\cal T})$ for $\tau=8.75$ and $\mathcal{T}/Q_0=(0.4,1,10)$. The horizontal dashed line indicates the  threshold $\theta$ (here fixed at $10^{-3}$). (b) Estimates of $\mu_{\Sigma}(1,{\cal T})$ versus time for different thresholds. Solid lines are a guide to the eye. The plateau   disappears as  $\theta$ is increased from $10^{-5}$  to $10^{-2}$.
}
\end{center}
\end{figure}
Consider for instance the data  for $\tau=8.75$ in Fig.~\ref{fig:SJ}d which shows a plateau  around $\mu_{\Sigma}(1,{\cal T}) \approx 0.024$ for $\theta=10^{-5}$.  As  seen in Fig.~\ref{Fig: convergence}b,  this plateau  shortens and eventually disappears as  $\theta$   increases from $10^{-5}$  to $10^{-2}$,  and fewer events contributing to the tails of $P(\Sigma_{{\cal T}})$  are taken into account. This behavior strongly suggests that the decrease of $\mu_{\Sigma}(1,{\cal T})$ displayed in Figs.~\ref{fig:SJ}c,d is a statistical artifact that would disappear \er{if all rare events were properly sampled} (a similar problem occurs with the  heat ${\cal Q}_{\cal T}$ in regime I, and the  {exact} fluctuation relation $\langle e^{-\beta {\cal Q}_{\cal T}}\rangle=e^{-{\cal T}/Q_0}$ cannot be verified for ${\cal T}/Q_0\gtrsim 2$  even with $5.10^6$ trajectories~\cite{SI}). To estimate $\mu_{\Sigma}(1)$, we thus  only consider the (reliable) ascending part of  $\mu_{\Sigma}(1,{\cal T})$, using the same fit as before \er{(a check of this procedure in presented in the Supplemental Material~\cite{SI})}. This yields very good agreement with the theoretical expression of  $\dot S_J$ for all delays  (Fig.~\ref{fig:SJ}e), from which  we conclude that the  conjecture for the asymptotic fluctuation relation  (\ref{EqIFT}) is  supported by experiments in regime II as well.
  
\textit{Conclusion.} We have experimentally demonstrated  the hidden role played by the acausality of the backward process associated with a non-Markovian feedback control. Even though the time-reversed dynamics is not physically realizable, it provides the proper tools (such as the acausal Jacobian or the acausal temperatures) to elucidate the nonequilibrium fluctuations of thermodynamic observables. The verification of these results  has required a careful statistical analysis of the asymptotic properties of the exponential average of work and entropy production, including their (nonexponential) prefactors, in order to obtain detailed information about the rare fluctuations of these observables. 
Our study opens up the exciting possibility of exploring  complex behaviors that are not easily amenable to  theoretical analysis, for instance those induced  by non-linearities in the observables or in the measurement and feedback protocol~\cite{A2021}.

\begin{acknowledgments}
N. K. acknowledges support from the Austrian Science Fund (FWF): Y 952-N36, START. We further acknowledge financial support from the German Science Foundation (DFG) under project FOR 2724.
\end{acknowledgments}

\clearpage
\newpage

\begin{center}
	\textbf{\large Supplementary Information - Non-Markovian feedback control and acausality: an experimental study}
\end{center}
\setcounter{equation}{0}
\setcounter{figure}{0}
\setcounter{table}{0}
\setcounter{page}{1}
\makeatletter
\renewcommand{\figurename}{Supplementary Figure}
\renewcommand{\theequation}{S\arabic{equation}}
\renewcommand{\thefigure}{S\arabic{figure}}
\renewcommand{\bibnumfmt}[1]{[S#1]}
\renewcommand{\citenumfont}[1]{S#1}

\title{Supplemental Material: Non-Markovian feedback control and acausality: an experimental study}
\author{Maxime Debiossac}
\email{maxime.debiossac@univie.ac.at}
\affiliation{University of Vienna, Faculty of Physics, VCQ, Boltzmanngasse 5, A-1090 Vienna, Austria} \author{Martin Luc Rosinberg}
\affiliation{LPTMC, CNRS-UMR 7600, Sorbonne Universit\' e, 4 place Jussieu, 75252 Paris Cedex 05, France}
\author{Eric Lutz}
\affiliation{Institute for Theoretical Physics I, University of Stuttgart, D-70550 Stuttgart, Germany}
\author{Nikolai Kiesel}
\affiliation{University of Vienna, Faculty of Physics, VCQ, Boltzmanngasse 5, A-1090 Vienna, Austria}
\maketitle

\section{Expression of the  stochastic entropy production $\Sigma_{{\cal T}}$}

As mentioned in the main text, the  time-integrated observable  $\Sigma_{{\cal T}}$ would be interpreted as the  total  trajectory-dependent entropy production by an observer unaware of the existence of the feedback loop~\cite{S_MR2014}. According to  Ref.~\cite{S_S2005},  this quantity  is defined as
\begin{align}
\label{EqSigmaTa}
\Sigma_{{\cal T}}&=\Sigma_{\cal T}^m+\Sigma_{\cal T}^{sys}\ ,
\end{align} 
where $\Sigma_{\cal T}^m=\beta {\cal Q}_{\cal T}$ is the change in the entropy of the medium and $\Sigma_{\cal T}^{sys}=\ln p(x_0,v_0)/p(x_{{\cal T}},v_{{\cal T}})$ is the change in the Shannon entropy of  the system (cf. Eq.~(4) in the main text). The heat $ {\cal Q}_{\cal T}$ is related to the work ${\cal W}_{\cal T}$  via the first law that expresses the conservation of energy at the microscopic level 
\begin{align}
\label{EqSigmaTb}
\beta {\cal Q}_{\cal T}=\beta{\cal W}_{{\cal T}}-\Delta {\cal U}(x_0,v_0,x_{\cal T},v_{\cal T})\ ,
\end{align} 
where
\begin{align}
\Delta {\cal U}(x_0,v_0,x_{\cal T},v_{\cal T})=\frac{1}{Q_0}[(x_{\cal T}^2-x_0^2)+(v_{\cal T}^2-v_0^2)]
\end{align} 
is the change in the internal energy of the system after time ${\cal T}$ (here expressed in reduced units). Moreover, since all forces acting on the particle are  linear, the stationary  distribution $p(x,v)$ is  a  bivariate Gaussian, 
\begin{align}
p(x,v)\propto e^{{-\frac{1}{2}\left [\frac{x^2}{\langle x^2\rangle}+\frac{v^2}{\langle v^2\rangle}\right]}}\ ,
\end{align}
as was checked experimentally in Ref. \cite{S_DGALK2020}. Replacing the mean-square position and the mean-square velocity by the configurational and kinetic temperatures, defined respectively as $T_x/T=(2/Q_0)\langle x^2\rangle$ and $T_v/T=(2/Q_0)\langle v^2\rangle$ in reduced units, we arrive at
\begin{align}
\label{EqSigmaTc}
\Sigma_{{\cal T}}&=\beta{\cal W}_{{\cal T}}+\frac{1}{Q_0} \left[\frac{T-T_x}{T_x}(x^2_{{\cal T}}-x_0^2)+\frac{T-T_v}{T_v}(v^2_{{\cal T}}-v_0^2)\right] \ ,
\end{align} 
where $\beta{\cal W}_{{\cal T}}$ is given by Eq.~(3) in the main text.

\section{Bifurcations and multistability}

\begin{figure}[t]
	\begin{center}
		\includegraphics[width=0.99\linewidth]{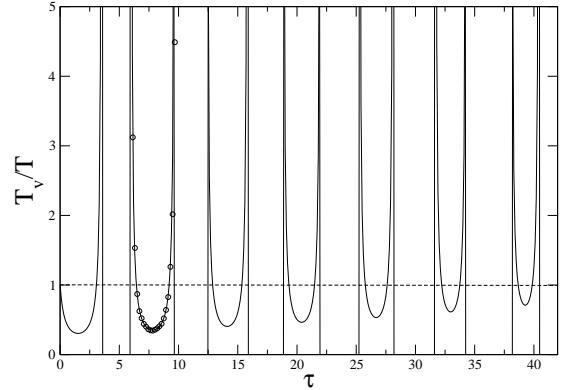}
		\caption{\label{FigB}  Kinetic temperature $T_v/T$ of the feedback-controlled oscillator as a function of  $\tau$: $T_v$  is positive in the regions where a stationary state exists (delimited by the vertical lines).  Only the first $7$ stability regions are shown. Experiments (circles) are performed in the second stability region for $5.996<\tau<9.658$.  There is no stationary state for a delay $\tau>89.628$.}
	\end{center}
\end{figure}

As shown in \el{Ref.~}\cite{S_RMT2015}, the  time-delayed harmonic oscillator obeying Eq.~(2)  in the main text  has a complex dynamical behavior. In particular, for $Q_0>\frac{1}{\sqrt{2}}$ and $\frac{1}{Q_0}\sqrt{1-\frac{1}{4Q_0^2}}<\vert \frac{g}{Q_0}\vert <1$, the oscillator features multistability with a characteristic "Christmas tree'' stability diagram. In the case under study, with $Q_0=50.2$ and $g=2.4$, there is an increasing sequence of critical delays ordered as follows,
\begin{align}
\label{Eqbifur}
\tau^*_{1,1}<\tau^*_{1,2}<\tau^*_{2,1}<...<\tau^*_{14,1}<\tau^*_{14,2}<\tau^*_{15,1}
\end{align}
with $\tau^*_{1,1}=3.506$, $\tau^*_{1,2}=5.996$, $\tau^*_{2,1}=9.658$,..., $\tau^*_{14,2}=89.521$, $\tau^*_{15,1}=89.628$ (in units $\Omega_0^{-1}$).  As $\tau$  varies  from $0$ to $\tau^*_{15,1}$, the  system switches from stability to instability and back to stability, and is unstable for  $\tau>\tau^*_{15,1}$.  A stationary solution only exists inside the stability domains and the temperatures $T_x$ and $T_v$ diverges at the boundaries, as illustrated in Fig.~\ref{FigB}. The present experimental study is performed in the second stability domain, for $\tau^*_{1,2}<\tau<\tau^*_{2,1}$.\\

\section{ Microscopic reversibility and acausal dynamics}

For the sake of completeness, we \el{summarize}   key points of the theoretical analysis performed in Refs.~\cite{S_MR2014,S_RMT2015,S_RTM2017}.

\subsection{Detailed fluctuation relation (or microscopic reversibility)}

In the presence of a time-delayed continuous feedback, the fluctuating heat exchanged between the Brownian particle and the thermal bath  along a trajectory ${\bf X}$  (see Fig.~1 in the main text) is not odd under time reversal. However, this property is preserved  if  the time-reversal operation is combined with the change \el{of the delay} $\tau \to -\tau$. The heat then obeys the modified detailed fluctuation relation~\cite{S_MR2014,S_RMT2015}
\begin{align}
\label{Eqpath0}
\beta {\cal Q}_{{\cal T}}=\ln \frac{{\cal P}[{\bf X}\vert  {\bf Y}]}{\widetilde {\cal P}[ {\bf X}^\dag\vert {\bf Y}^\dag]}+\ln \frac{\widetilde {\cal J}[{\bf X}]}{{\cal J}}\ ,
\end{align}
where  ${\cal P}[{\bf X}\vert  {\bf Y}]$ is the probability of  ${\bf X}$, given the path ${\bf Y}$ in the time interval $[-\tau,0]$, and   $\widetilde {\cal P}[ {\bf X}^\dag\vert {\bf Y}^\dag]$  is the probability of  ${\bf X}^\dag$, given the time-reversed path ${\bf Y}^\dag$. The  ``tilde" symbol refers to the  ``conjugate" acausal  dynamics defined by changing   $\tau$ into  $-\tau$.  In general, the  corresponding Jacobian $\widetilde {\cal J}[{\bf X}]$ of the transformation $\xi(t)\rightarrow x(t)$ [see Eq.~(2) in the main text] is a nontrivial functional of the path, in contrast with  the Jacobian ${\cal J}$ of the original causal dynamics. When the system settles in \el{a} stable stationary state, one can define the asymptotic rate 
\begin{align}
\label{EqdefRcg}
\dot {\cal S}_{\cal J}= \lim_{{\cal T}\rightarrow \infty}  \frac{1}{{\cal T}}\langle\ln \frac{\cal J}{\widetilde {\cal J}[{\bf X}]}\rangle_\text{st}\ ,
\end{align}
which  turns out to be an upper bound to the extracted work rate $\dot {\cal W}_\text{ext}=-\dot {\cal W}$, i.e., 
\begin{align}
\label{Eq2law}
\beta \langle \dot {\cal W}_\text{ext}\rangle_\text{st} \le \dot {\cal S}_{\cal J}\ .
\end{align}
Note that $\dot {\cal S}_{\cal J}$ generally differs from the so-called "entropy pumping rate" $\dot {\cal S}_\text{pump}$, which  is also an upper bound to the extracted work, as tested experimentally in \el{Ref.~}\cite{S_DGALK2020}. 

The  second-law-like inequality (\ref{Eq2law}) is a consequence of the integral fluctuation theorem $\langle e^{-R_{cg}[{\bf X}]}\rangle=1 $ where $R_{cg}[{\bf X}]$ is a rather complicated path-dependent quantity that is not accessible to experiments~\cite{S_RMT2015}. However, one has $\dot {\cal R}_{cg}=\beta \dot {\cal Q}+\dot {\cal S}_{\cal J}$  in the long-time limit, where $\dot {\cal R}_{cg}= \lim_{{\cal T}\rightarrow \infty}\frac{1}{\cal T}\langle R_{cg}[{\bf X}]\rangle_{st}$, and the inequality  $\langle R_{cg}[{\bf X}]\rangle\ge 0$ implies  inequality (\ref{Eq2law}). 

In the case of a linear Langevin dynamics,  $\dot {\cal S}_{\cal J}$ can  be explicitly expressed  in terms of the poles of the acausal response function $\widetilde \chi(s)\equiv \chi(s)_{\tau \to -\tau}$ in the complex Laplace plane, where $\chi(s)$ is the standard response function of the oscillator in the Laplace representation~\cite{S_RMT2015}. Specifically, 
\begin{align}
\label{Eqchis}
\widetilde \chi(s)=[s^2+\frac{s}{Q_0}+1-\frac{g}{Q_0}e^{s\tau}]^{-1} 
\end{align}
in reduced units. $\dot {\cal S}_{\cal J}$ is then given by Eq.~(160) in Ref.~\cite{S_RMT2015}.

\subsection{Acausal dynamics}

Although the acausal dynamics defined by the Langevin equation (here written in reduced units)
\begin{align}
\ddot{x}_t+ \frac{1}{Q_0}\dot{x}_t+ x_t -\frac{g}{Q_0} x_{t+\tau}=\xi_t,
\label{EqLacausal}
\end{align} 
is not physically realizable, a stationary solution may still exist, characterized as usual by the fact that the $n$-point probability distributions are invariant under time translation. This also means that the solution  must be independent of both the initial condition in the far past and the final condition in the far future, such that
\begin{align}
x(t)\approx \int_{-\infty}^{+\infty} dt'\: \widetilde \chi(t-t')\xi(t')\ ,
\label{Eqacausal1}
\end{align}
where $\widetilde \chi(t)$ is the acausal response function in the time domain. This implies that $\widetilde \chi(t)$  decreases sufficiently  fast for  $t\to \pm\infty$  and in this case $\widetilde \chi(t)$ is  just  the inverse Fourier transform of $\widetilde \chi(\omega=is)$ and \el{vice versa} (in general, $\widetilde \chi(t)$ is defined as  the inverse bilateral Laplace transform of the function $\widetilde \chi(s)$ defined by Eq. (\ref{Eqchis}) and it may have no Fourier transform~\cite{S_RTM2017}).   Concretely, this requires  that  $\widetilde \chi(s)$ has two and only two poles on the l.h.s. of the complex $s$-plane. For the case under study, with the system operating in the second stability region, this occurs for $5.996< \tau<6.854$ and $8.797< \tau<9.658$ (with $\Omega_0^{-1}$ as the time unit). The   configurational and kinetic temperatures of the ``acausal" stationary state (the latter \el{quantity} plotted in  Fig.~3 of the main text) are then obtained from 
\begin{align}
\frac{\widetilde T_x}{T}&=\frac{2}{Q_0} \int_{-\infty}^{+\infty} \frac{d\omega}{2\pi}\vert \widetilde \chi(\omega)\vert^2\nonumber\\
\frac{\widetilde T_v}{T}&=\frac{2}{Q_0} \int_{-\infty}^{+\infty} \frac{d\omega}{2\pi}\omega^2\vert \widetilde \chi(\omega)\vert^2\ .
\label{Eqacausal2}
\end{align}

\subsection{Asymptotic behavior of $\langle e^{-\Sigma_{{\cal T}}}\rangle$ and $\langle e^{-\beta {\cal W}_{{\cal T}}}\rangle$}

When the acausal Langevin Eq.~(\ref{EqLacausal}) admits a stationary solution,  the theoretical analysis of the  asymptotic behavior of the  generating functions $Z_{\Sigma}(\lambda, {\cal T})=\langle e^{-\lambda \Sigma_{{\cal T}}}\rangle$  and $Z_{W}(\lambda, {\cal T})=\langle e^{-\lambda \beta {\cal W}_{{\cal T}}}\rangle$ for ${\cal T} \to \infty$ and $\lambda=1$  \el{suggests} that $\mu_W(1)= \mu_{\Sigma}(1)=\dot {\cal S}_{\cal J}$~\cite{S_RTM2017}. 
\el{At the same time}, the expression of the ratio $g_W(1)/g_{\Sigma}(1)$ of the pre-exponential factors (Eq.~(4) in the main text) is more an educated guess based on  perturbative calculations and on the exact behavior of the system in some limiting cases (e.g., the small-$\tau$ and large-$Q_0$ limits). 

\el{On the other hand,} when the "acausal"  stationary solution  does not exist, things are more complicated. However, the explicit result $\mu_W(1)=1/Q_0$ can be proven when the conjugate dynamics  obtained by changing the sign of the friction coefficient in the original Langevin equation (the so-called  ``hat" dynamics in \el{Ref.}~\cite{S_RTM2017}) relaxes to a stationary state.
This is  what occurs for $6.85\lesssim\tau\lesssim 8.80$ (grey region in Fig.~4e of the main text).
On the other hand, there is yet no exact theoretical prediction for $\mu_{\Sigma}(1)$, and the main conjecture tested in this work, stating that $\mu_{\Sigma}(1)$ is {\it always} equal to $\dot {\cal S}_{\cal J}$,  is  based on a limited set of simulation results.  

\subsection{Stationary-state fluctuation theorem}

Because of the non-Markovian feedback,  the system does not obey a conventional stationary-state fluctuation theorem (SSFT) 
expressing the symmetry around $0$ of the pdf of an observable  at large times, for instance $\lim_{{\cal T} \to \infty} \frac{1}{{\cal T}}\ln [P(\beta {\cal W}_{\cal T}=w{\cal T})/{P(\beta {\cal W}_{\cal T}=-w{\cal T})}]= w$ for the work done by an external force (see e.g. Ref.~\cite{S_C2017}). However, provided the acausal  stationary state exists, this relation is replaced by~\cite{S_RTM2017} 
\begin{align}
\label{EqSSFT}
\lim_{{\cal T} \to \infty} \frac{1}{{\cal T}} \ln \frac{P(\beta {\cal W}=w{\cal T})}{\widetilde P(\beta \widetilde{\cal W}=-w {\cal T})}=w+\dot S_{\cal J}\ ,
\end{align}
where  $\beta \widetilde{\cal W}_{\cal T}\equiv \beta {\cal W}_{\cal T}\vert_{\tau \to -\tau}=(2g)/(Q_0^2)\int_0^{{\cal T}} dt'\: x_{t'+\tau}\circ v_{t'}$. 

\begin{figure}[t]
	\begin{center}
		\includegraphics[width=0.99\linewidth]{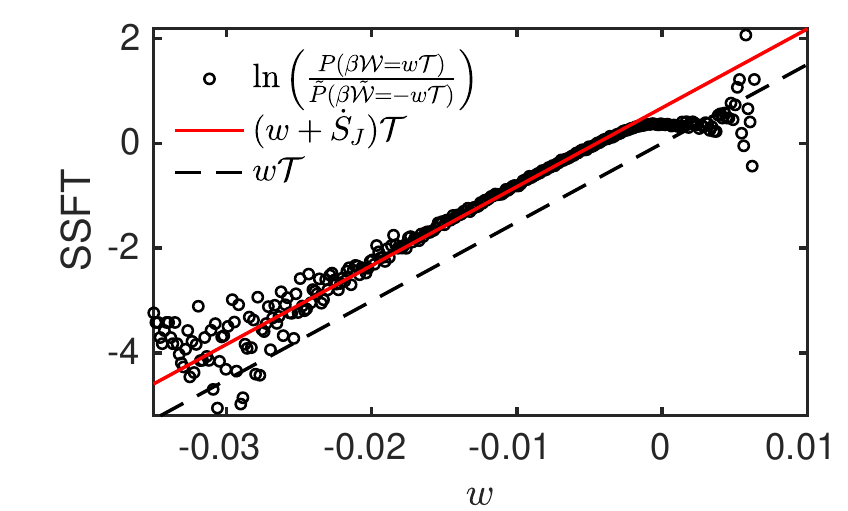}
		\caption{\label{FigSSFT}  Verification of the modified \el{stationary-state fluctuation theorem} (SSFT) $\ln [P(\beta {\cal W}=w{\cal T})/\widetilde P(\beta \widetilde{\cal W}=-w {\cal T})]\sim(w+\dot{S}_J){\cal T}$ for $\tau=6.51$ ($\Omega_0^{-1}$ is the time unit). The trajectory length is ${\cal T}=3Q_0$. The ``acausal" probability $\widetilde{P}(\beta \widetilde{\cal{W}}=-w{\cal T})$  is obtained as explained in the text.}
	\end{center}
\end{figure}

A check of this modified SSFT is shown in Fig.~\ref{FigSSFT} for  $\tau=6.51$, with $P(\beta {\cal W}_{{\cal T}}=w{\cal T})$ and $\dot S_J $  obtained  from experiments (specifically, $\dot S_J\approx 0.0045$ from the fit of $\mu_\Sigma(1,\mathcal{T})$, which is close to the theoretical value, as can be seen in Fig. 4e of the main text). It is obviously crucial to include this contribution  in the SSFT  to reach a good agreement. To compute the ``acausal" probability $\widetilde P(\beta \widetilde{\cal W}=-w {\cal T})$, a large number of stationary trajectories  of length ${\cal T}$ generated by  Eq.~(\ref{EqLacausal}) is needed. Like in \el{Ref.~}\cite{S_RTM2017}, this can done  by computing numerically the  response function $\widetilde \chi(t)$ and then using Eq.~(\ref{Eqacausal1}) (the  case shown in Fig.~\ref{FigSSFT} turns out to be rather  challenging because  $\widetilde \chi(t)$ decreases very slowly with $t$ for $t>0$). Another method  is to solve the acausal Langevin equation (\ref{EqLacausal}) iteratively as
\begin{align}
\dot{v}_t^{(n)}=-\frac{1}{Q_0}v_t^{(n)}-x_t^{(n)}+\frac{g}{Q_0}x_{t+\tau}^{(n-1)}+\xi_t\ ,
\label{EqLiter}
\end{align}
starting for some  initial trajectory $\{x_t^{(0)}\}$ of length  ${\cal T}^{(0)}\gg{\cal T}$ and  reducing the length of the trajectory by  $\tau$ at each iteration. This procedure  circumvents the obstacle  of non-causality since an entire time sequence  is available from the  previous iteration. This is somewhat similar to the non-causal learning algorithm used in iterative learning~\cite{S_BTA2006}.  However, convergence of the procedure is not guaranteed and  requires to properly choose the initial trajectory.  Of course, this cannot be done experimentally.

\section{Method and statistical analysis}

\subsection{Block averaging and rare events}

The estimates of the scaled cumulant generating functions (SCGF) $\mu_W(1)$ and $\mu_{\Sigma}(1)$ are obtained by using a  block averaging approach. In the context of large deviation theory, this method is well-suited  for integrated observables of continuous-time  random processes~\cite{S_DM2005, S_RAT2015,S_RB2020}. We thus divide the recorded trajectory of length ${\cal T}_{tot}=1000$ s into $N$ blocks (i.e., trajectories) of length ${\cal T}$, separated one from another by $\Delta {\cal T}$. Hence ${\cal T}_{tot}=N{\cal T}+(N-1)\Delta {\cal T}$. To ensure that $x(t)$ and $x(t-\tau)$ belong to the same block,  $\Delta {\cal T}$ is much larger than the largest value of $\tau$ considered in our experiments.  Specifically, $\Delta {\cal T}\approx 0.053$ ms (i.e. $\Delta {\cal T}=2Q_0=100.4$  in reduced units whereas $\tau_{max} \approx 9.50$).  The observables are then computed in each block. 
Consider for instance the work. The estimator of $Z_W(1,{\cal T})= \langle e^{-\beta {\cal W}_{{\cal T}}}\rangle$  is given by
	\begin{align}
	\hat Z_W( 1,N,{\cal T})=\frac{1}{N}\sum_{j=1}^Ne^{-\beta {\cal W}_{{\cal T}}^{(j)}}\ ,
	\end{align}
	where $\beta W_{{\cal T}}^{(j)}=\frac{2g}{Q_0^2}\int_{t_j}^{tj+{\cal T}}dt \: x_{t-\tau}\circ v_{t}$ and  $t_j=(j-1)({\cal T}+\Delta {\cal T})$ is the initial time of block $j$. Since  time is discretized,  the velocity is computed  as $v_i=(x_{i+1}-x_i)/\Delta t$ (with $\Delta t=0.2 \mu$s),  and  a spline interpolation of the time series is used to precisely pinpoint the position of the particle at time $t-\tau$.

The estimator of  $\mu_W(1,{\cal T})\equiv(1/{\cal T} )\ln Z_W( 1,{\cal T})$ is  then 
	\begin{align}
	\hat \mu_W( 1,N,{\cal T})=\frac{1}{\cal T}\ln  \hat Z_W( 1,N,{\cal T})\ .
	\end{align}
	In principle, two limits must be  successively taken to obtain the SCGF: first $\lim_{N \to \infty}\hat \mu_W( 1,N,{\cal T})=\mu_W(1,{\cal T})$, and then  $\lim_{{\cal T} \to \infty}\mu_W( 1,{\cal T})=\mu_W(1)$. In practice, however, one is constrained by the fixed length ${\cal T}_{tot}$ of the recorded time series,  and one faces a trade-off in the choice of ${\cal T}$  and $N$~\cite{S_RB2020}.  
	For instance,  one has  ${\cal T}\le {\cal T}_{\max}\approx 0.15$  ms with $N=5. 10^6$ blocks and ${\cal T}\le {\cal T}_{\max}\approx  0.29$ ms  with $N=3.10^6$ (respectively ${\cal T}_{\max}/Q_0\approx 5.45$ and ${\cal T}_{\max}/Q_0\approx 10.42$ in reduced units). 
	In fact, Figs. 4 and 5 in the main text show that the rare events  which make $\langle e^{-\Sigma_{\cal T}}\rangle$  different from  $\langle e^{-{\cal W}_{\cal T}}\rangle$ in regime II come into play for ${\cal T}$ significantly smaller than  ${\cal T}_{\max}$. 
\begin{figure}[t]
	\begin{center}
		\includegraphics[width=0.99\linewidth]{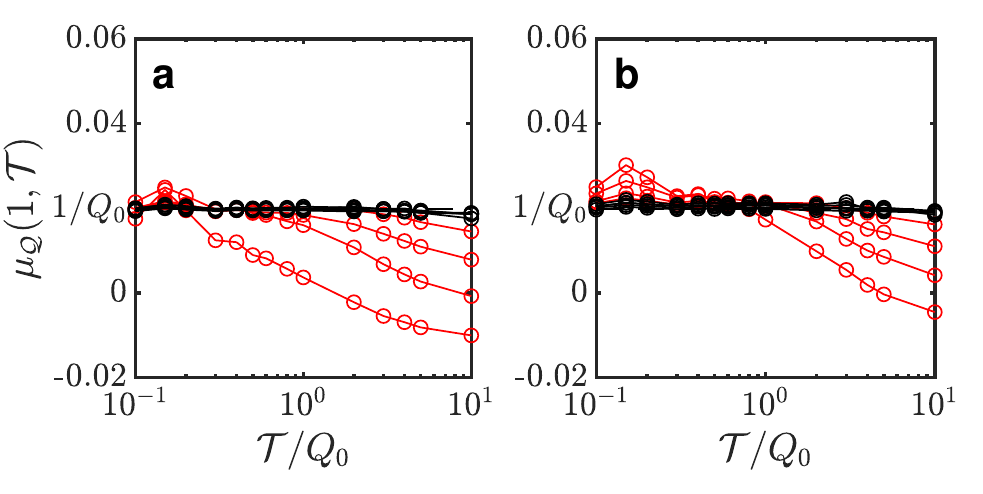}
		\caption{\label{Figmuheat1} Estimates of $\mu_Q(1,{\cal T})$ as a function of the observation time ${\cal T}$ for delays $6.13\leq\tau\leq 7.82$ (a) and $8.00\leq\tau\leq 9.50$ (b). The values of the delay $\tau$ and the color of the symbols are the same as in Fig.~4 of the main text.}
	\end{center}
\end{figure}

Remarkably,  the dissipated heat ${\cal Q}_{\cal T}$ displays a similar behavior in regime I. The crucial difference with the entropy production is that  ${\cal Q}_{\cal T}$ satisfies  the integral fluctuation theorem (IFT)  $\langle e^{-\beta {\cal Q}_{\cal T}}\rangle=e^{-(\Gamma/m) {\cal T}}$  at {\it all} times (a universal result for underdamped Langevin dynamics~\cite{S_RTM2016}). Hence,  one should observe that $\mu_Q(1,{\cal T})=1/Q_0$ in reduced units. However, as shown in Fig.~\ref{Figmuheat1}, this is only observed in regime II (black symbols in the figure),  that is, when the "hat" dynamics~\cite{S_RTM2017} obtained by changing the sign of the friction coefficient  relaxes to a stationary state. In regime I (red symbols), the estimated values of $\mu_Q(1,{\cal T})$ exhibit a spurious behavior as a function of ${\cal T}$ which resembles the one found for $\mu_{\Sigma}(1,{\cal T})$. 
\begin{figure}[t]
	\begin{center}
		\includegraphics[width=0.8\linewidth]{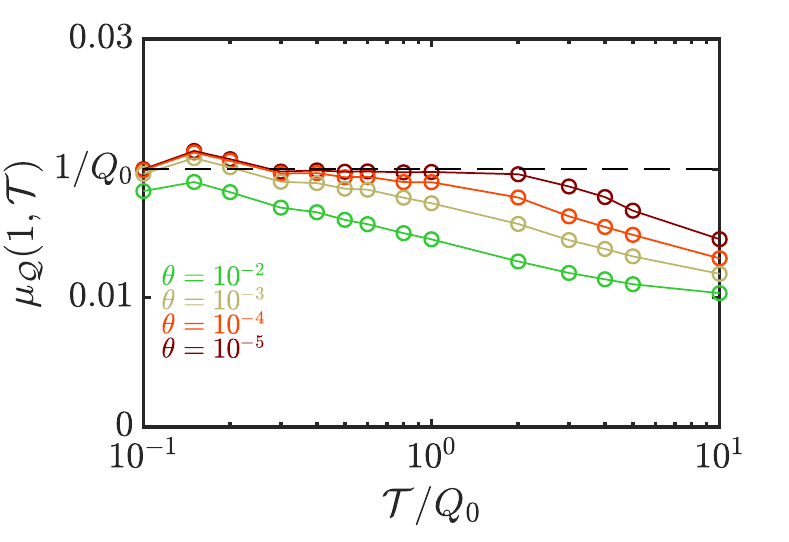}
		\caption{\label{Figmuheat2} Estimates of $\mu_Q(1,{\cal T})$ versus time for  $\tau=6.70$ and different thresholds. Dashed lines are a guide to the eye. One should observe $\mu_Q(1,{\cal T})=1/Q_0$  but  the plateau disappears as $\theta$ is increased from $10^{-5}$ to $10^{-2}$.  }
	\end{center}
\end{figure}

Figure \ref{Figmuheat2} (for $\tau=6.70$)  is similar to Fig. 5b in the main text and clearly  shows that  the correct IFT  is progressively recovered (at least for ${\cal T}/Q_0 \lesssim 2$) as  the threshold $\theta$  is decreased and more events contributing to the tails of the pdf  of the heat are included in the calculation of the exponential average. This  further supports our claim that the fall-off of $\mu_{\Sigma}(1,{\cal T})$ for ${\cal T}/Q_0\gtrsim 1$ is also spurious.

\subsection{Extrapolation and errors}

\begin{figure}[b]
	\begin{center}
		\includegraphics[width=1\linewidth]{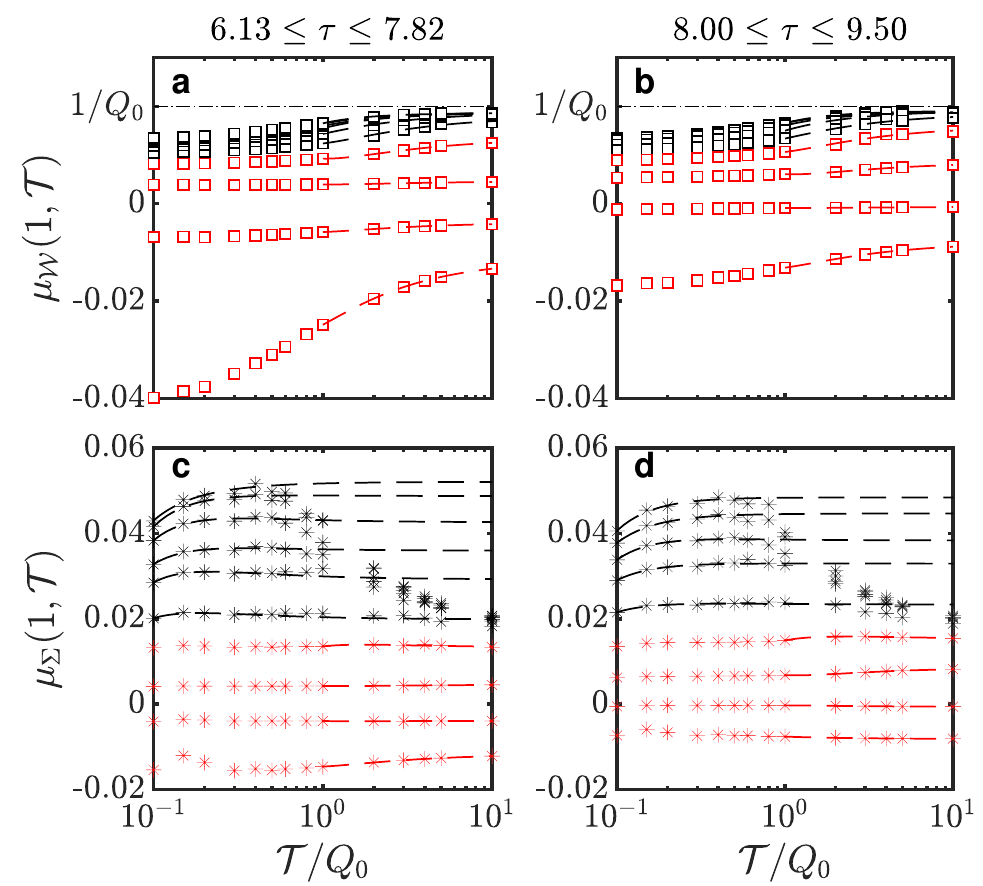}
		\caption{\label{FigFit} Same as Figs.~4a-d in the main text with the dashed lines representing the fits of the finite-time values that are used to extract the asymptotic scaled cumulant generating functions and the pre-exponential factors in regime I. 
		}
	\end{center}
\end{figure}

\begin{figure}[t]
	\begin{center}
		\includegraphics[width=0.9\linewidth]{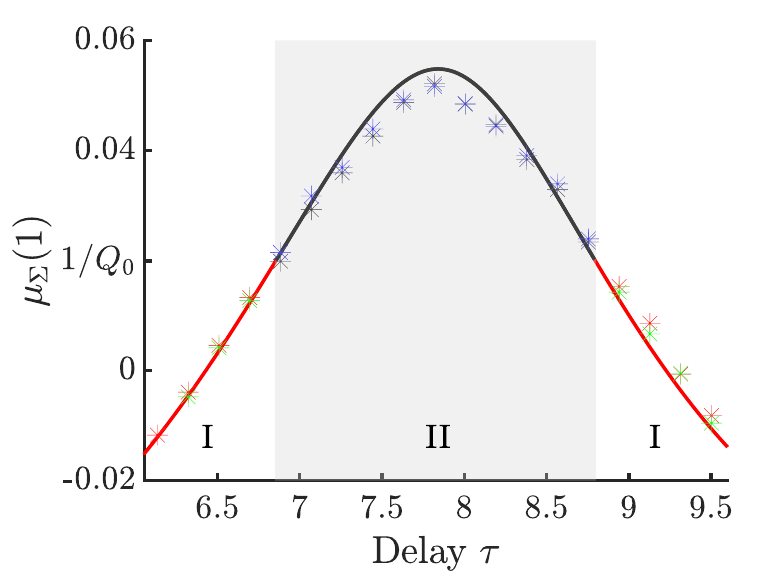}
		\caption{\label{FigFit2} Comparison of the SCGF $\mu_\Sigma(1)$ using different analysis methods. In regime I, red stars results from a fit performed in the interval $1\leq\mathcal{T}/Q_0\leq 10$, while green stars is for a fit in the interval $0.1\lesssim {\cal T}/Q_0\lesssim 0.4$. Note that, for $\tau=6.13$ (first delay), this latter fit cannot be done because of the non-monotonic behavior of $\mu_{\Sigma}(1,{\cal T})$ at small times. In regime II, black stars are obtained with a fit in the interval $0.1\lesssim {\cal T}/Q_0\lesssim 0.4$ and are compared with the values given by the maximum of $\mu_{\Sigma}(1,{\cal T})$ (blue stars).}
	\end{center}
\end{figure}

The infinite-time limit of the estimators of the SCGF's  are extracted from the finite-time results  using a  fit of the form  $a+b/{\cal T}+c/{\cal T}^2$.  We note that this type of $1/{\cal T}$ scaling has also  been used to evaluate large-deviation functions obtained by a population dynamics algorithm~\cite{S_HNL2017}.     This leads to the fits  represented by the dashed lines in Fig. \ref{FigFit}.  
We recall that the fits in regime I (red dashed lines) are done in the range  $1\le {\cal T}/Q_0\le 10$ for both the work and the entropy production. This allows us to obtain  a sensible estimation of the pre-exponentials factors $g_W(1)$ and $g_{\Sigma}(1)$ in addition to the asymptotic SCGFs. On the other hand,   the values of $\mu_{\Sigma}(1,{\cal T})$  in regime II (black dashed lines) are fitted for  ${\cal T}/Q_0\lesssim 0.4$ only since the statistics of the rare events is clearly deficient at larger times.  As shown in Fig. \ref{FigFit2} (which corresponds to Fig. 4e in the main text),  the fit of $\mu_{\Sigma}(1,{\cal T})$  in regime I can also be done for ${\cal T}/Q_0\lesssim 0.4$ without significantly changing the extrapolated values of the SCGF.  However,  one can no longer extract a reliable value of the pre-exponential factor $g_{\Sigma}(1)$  from the  $1/{\cal T}$ correction.


\begin{figure}[t]
	\begin{center}
		\includegraphics[width=1\linewidth]{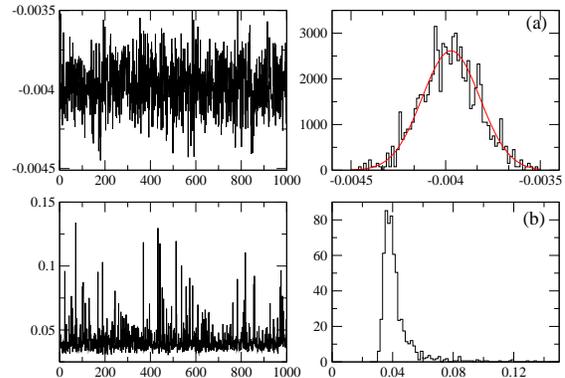}
		\caption{\label{FigErrors}  Values of $\langle \hat \mu_{\Sigma}( 1,N,{\cal T})\rangle_b$ calculated in $1000$ blocks of size $n=5000$ and  the corresponding statistical distributions for ${\cal T}/Q_0=1$, $\tau =6.32$ (a) and $\tau=7.82$ (b).  The solid red line in panel (a) is the best fit by a Gaussian distribution. The distribution is non-Gaussian in panel (b).}
	\end{center}
\end{figure}

It would be desirable to provide confidence intervals for  $\mu_{W} (1,{\cal T})$, $ \mu_{\Sigma}(1,{\cal T})$,  and  eventually for the SCGFs. This requires to estimate the statistical and systematic errors due to the finite value of $N$, that is the variance and the bias  of the estimators (see e.g., Refs.\cite{S_ZW2002,S_GRB2003,S_PJC2010}), in addition to the experimental uncertainties in the values of the damping rate  and the feedback gain  mentioned in the main text. However,  this is a very challenging task in the present case because the probability distributions of the observables are not Gaussian. In particular, $P(\Sigma_{\cal T})$ has an exponential tail on the left-hand side, which in the large-deviation regime is associated with the presence of poles in the pre-exponential factor~\cite{S_RTM2017}.  The main problem is that the distribution of $e^{-\Sigma_{\cal T}}$ may not have a finite second moment, so  that  the estimator  may  not converge asymptotically to a Gaussian distribution around the mean~\cite{S_RAT2015}. This  crucially depends on the value of $\tau$, as   illustrated in  Fig. \ref{FigErrors} for ${\cal T}/Q_0=1$ and two different values of $\tau$ (respectively in regime I and II).  The statistical distributions have been obtained  by grouping the $N=5.10^6$ trajectories  into  $N_b=1000$ blocks of size $n=N/N_b=5000$,  and  computing the average $\langle \hat \mu_{\Sigma}( 1,N,{\cal T})\rangle_b=(1/{\cal T})\langle \ln \hat Z_{\Sigma}( 1,N,{\cal T})\rangle_b$  in each block. For $\tau=6.32$,  the variance of $P(e^{-\Sigma_{\cal T}})$ is finite, and it can be seen that the  distribution of $\langle \hat \mu_{\Sigma}( 1,N,{\cal T})\rangle_b$ is reasonably Gaussian. On the other hand, the variance  diverges for $\tau=7.82$, and  the  distribution of $\langle \hat \mu_{\Sigma}( 1,N,{\cal T})\rangle_b$ is strongly asymmetric with a long tail on the right-hand side. In the first case, the statistical error is the dominant source of noise~\cite{S_GRB2003,S_PJC2010}, but the final error on $\hat \mu_{\Sigma}(1,{\cal T})$  turns  out to be very small (of the order $10^{-6}$, which is  too small to be visible in Fig. 4 in the main text).    In the second case, it is very likely that the bias due to the undersampling of the rare events associated with the temporal boundary term  is dominant but, unfortunately, a reliable estimate  cannot be obtained. This difficult issue is left for future investigations.


\begin{thebibliography}{10}
\bibitem{note5} For simplicity, we use the same notation for the dimensional and dimensionless parameters.
\end{thebibliography}
\end{document}